\documentclass[aps,prb,twocolumn,floatfix,superscriptaddress]{revtex4}
\usepackage{graphicx}
\usepackage{amssymb}
\usepackage{amsfonts}
\usepackage{amsmath}
\usepackage{dcolumn}
\usepackage{soul}
\usepackage{makecell}
\usepackage{color}
\usepackage{stackrel}
\usepackage{latexsym}
\usepackage[normalem]{ulem}
\usepackage{times}
\usepackage{bm}
\usepackage{hyperref}
\usepackage{accents}
\usepackage{rotating}
\usepackage{adjustbox}
\usepackage{chngcntr}

\definecolor{capri}{rgb}{0.0, 0.75, 1.0}

\begin{document}

\title{Correlation-temperature phase diagram of prototypical infinite layer rare earth nickelates}

\author{Gheorghe Lucian Pascut}
\affiliation{MANSiD Research Center and Faculty of Forestry, Stefan Cel Mare University (USV), Suceava 720229, Romania}

\author{Lucian Cosovanu}
\affiliation{MANSiD Research Center, Stefan Cel Mare University (USV), Suceava 720229, Romania}

\author{Kristjan Haule}
\affiliation{Department of Physics \& Astronomy, Rutgers University, Piscataway, NJ O8854, USA}

\author{Khandker F. Quader}
\affiliation{Department of Physics, Kent State University, Kent , OH 44242, USA}

\date{\today}

\begin{abstract}
The discovery of superconductivity in hole-doped infinite layer nickelates, $R$NiO$_2$ ($R$ = Nd, Pr, La) has garnered sustained interest in the field. A definitive picture of low-energy many-body states has not yet emerged. We provide new insights into the low-energy physics, based on our embedded dynamical mean-field theory calculations, and propose a correlation (U)-temperature (T) phase diagram. The key features are a low-T Fermi liquid (FL) phase, a high-T Curie-Weiss regime, and an antiferromagnetic phase in a narrow U-T region. We associate the onset of the FL phase with partial screening of Ni-d moments; however, full screening occurs at lower temperatures. This may be related to insufficiency of conduction electrons to effectively screen the Ni-d moments, suggestive of Nozieres Exhaustion Principle. Our results suggest that $R$NiO$_2$ are in the paramagnetic state, close to an antiferromagnetic dome, making magnetic fluctuations feasible. This may be consequential for superconductivity.
\end{abstract} 

\maketitle
\clearpage
\newpage
\mbox{~}
\clearpage
\newpage

\section{Introduction}
\label{sec_Introduction}

The quest for discovery of  superconductivity at high temperatures has led to exploration of classes of materials other than the cuprates, resulting
in exciting developments in superconductivity, and the unveiling of other novel properties of quantum materials. The recent observation of superconductivity in hole-doped infinite layer 
nickelates, $R$NiO$_2$ ($R$ = rare earth Nd, Pr, La)~\cite{Li-Nature19,Osada-Pr-Nano20,Zeng-La-SciAdv} have generated tremendous interest in the field of condensed matter. Vigorous, innovative efforts in experiment and theory have been driven by the prospect of $R$NiO$_2$ constituting another class of high T$_C$ superconductors, with the possibility of unconventional pairing. There is also the expectation that 
the infinite-layer nickelates, being iso-structural to the cuprates, may provide another perspective on understanding high T$_C$ superconductivity in the cuprates. 

While magnetic excitations and damped spin waves have recently been observed in resonant inelastic x-ray scattering experiment in infinite layer NdNiO$_2$~\cite{Lu-magexcit-Science22}, long-range magnetic order has not been observed 
so far. Earlier neutron scattering experiments~\cite{Hayward99,Hayward03} on LaNiO$_2$ and NdNiO$_2$ did not find long-range magnetic order down to 5K and 1.7K respectively. Likewise, a combination of $\mu$sr, magnetic susceptibility and specific heat measurement~\cite{Ortiz-PRR22} showed no long-range magnetic order down to 2K in polycrystalline LaNiO$_2$.
Spin susceptibility measurements on LaNiO$_2$~\cite{Hayward99} showed weak temperature (T) dependence and pointed to two Curie-Weiss regimes, and recent experiment~\cite{Ortiz-PRR22} indicates deviation from Curie-Weiss behavior at high temperatures.
Specific heat measurement~\cite{Ortiz-PRR22} on the same compound finds a linear-T behavior, suggestive of Fermi liquid behavior at low temperature.

To be able to arrive at an understanding of the mechanism of superconductivity, it is important to understand the 
electronic structure, nature and strength of correlations, and excitations in the normal state of the nickelates. Besides several model calculations~\cite{Sawatzky-PRL20, Zhang-PRB20,Hu-PRR19,Vishwanath-model-PRR20,Hoshino-model-PRB20}, there have been a number of first principle
work, that generally fall into the categories of density functional theory (DFT and DFT + U)~\cite{Botana-Norman-DFT-PRX20,Nomura-DFT-PRB19,Moritz-DFT-PRX21,Anisimov-DFT-PRB99,Pickett-DFT-PRB04}.
and dynamical mean-field theory (DMFT).~\cite{Lechermann-PRB20,Gu-comphys20,Karp-PRX20,Wang-Kotliar-PRB20,Kang-Kotliar-Hunds-PRL21,Kitatani-DMFT-NatQM20,Savrasov-DMFT-PRB20}
There has also been work based on dynamical vertex approximation.~\cite{Held2022}
Since the nickelates are iso-electronic to the cuprates, there have been a number of useful efforts exploring the similarities and differences between the electronic structures of the two, namely,
 $d^8$ versus $d^9$ electronic configuration of Ni; charge transfer energy, $\epsilon_d - \epsilon_p$, compared to that in the cuprates, and possibility of magnetic or insulating behavior in the nickelates. 

\begin{figure*}[!htb]
\includegraphics[width=0.80\linewidth]{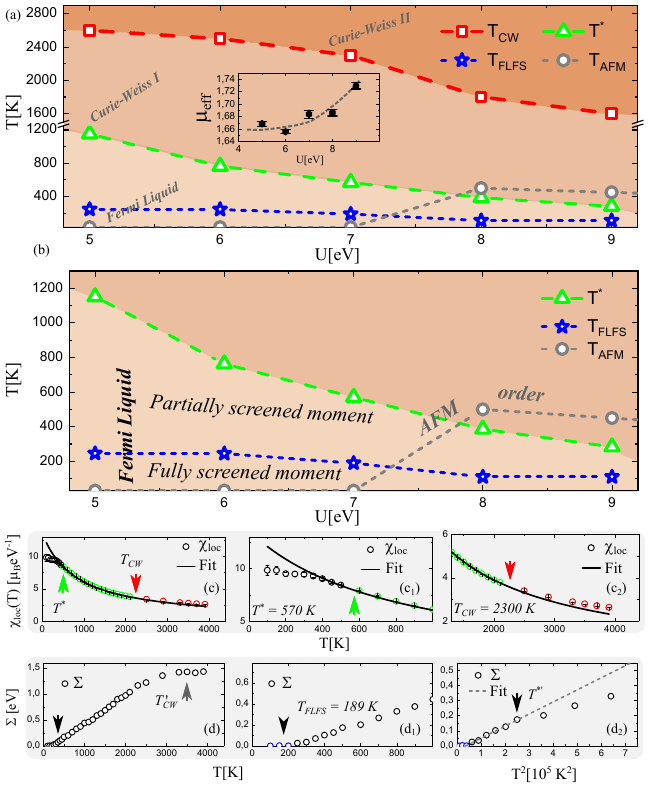}
\caption{(Color online) \textbf{U - T Phase diagram LaNiO$_2$:} Panel (a) shows the correlation-temperature (U - T) phase diagram where the boundaries of various electronic phases were determined from the fits of 
the calculated local magnetic susceptibility ($\chi_{loc}$) for the nickel d electrons, and scattering rate ($\Sigma$) for the electrons in the nickel $d_{x^2-y^2}$ orbital. The red ($T_{CW}$) and green ($T^*$) dashed lines denote the Curie-Weiss and Fermi liquid phase boundaries  The boundary (black dashed curve, $T_{AFM}$) of the antiferromagnetic (AFM) phase was obtained by computing the temperature dependence of the magnetic order parameter, described in the text. The inset shows the U dependence of the local effective moment $\mu^2_{eff}$, where the dashed line is a guide to the eye. Panel (b) shows the low-temperature region of the U - T phase diagram, showing the regions of partially screened and fully screened (below $T_{FLFS}$) Fermi liquid phases.
Panel (c) shows the temperature dependence of the calculated local magnetic susceptibility ($\chi_{loc}$) for the Ni-d electrons for U = 7 eV and J = 1 eV. Panels (c$_1$) and (c$_2$) show $\chi_{loc}$ with the focus on the low- and high- temperature ranges respectively. The solid line represents the best fit with the Curie-Weiss-Wilson formula $\chi_{loc} = \frac{C \mu^2_{eff}}{(T + 2 T_S)}$ as explained in the text. From panels (c$_1$) and (c$_2$) we observe that $T^*$ and $T_{CW}$ corresponds to the temperatures where the $\chi_{loc}$ shows deviations from the Curie-Weiss behavior.  Panel (d) shows the temperature dependence of the calculated scattering rate ($\Sigma$) for the electrons in the Ni $d_{x^2-y^2}$ orbital, for the same value of U and J used for the susceptibility. Panels (d$_1$) and (d$_2$) focus on the low-temperature region, where $\Sigma$  is plotted versus T and $T^2$ respectively. The dashed line represents the best linear fit with the formula $\Sigma$ = $A ( T^2 - T^{*2}) $, where A $>$ 0. $T_{CW}^{'}$, $T^{*'}$ and $T_{FLFS}$ are the temperatures at which $\Sigma$ flattens at high-temperature, the system starts to show a Fermi liquid behavior (linear dependence of $\Sigma$ on T$^2$), and $\Sigma$ goes to zero at low-temperature respectively.  Note that $T^{*'}$ $\approx$ $T^*$.  The arrows show the position of various critical temperatures.}
\label{fig:UT_PD}
\end{figure*}

It would be desirable to have a more definitive understanding of the the low-energy many-body states of the nickelates, in particular, magnetic versus paramagnetic, insulating versus metallic or localized Ni-d moment versus itinerant behavior, and if the physics is governed by single-orbital (as in the cuprates) or multi-orbital degrees of freedom. Likewise, a study that encompasses sufficiently wide ranges of temperature 
and correlation strength can shed light on the system behavior, phases and transitions between them. 

To understand the nature of the many-body states of the $R$NiO$_2$ class of materials at low, as well as, higher temperatures, we perform first principle embedded DMFT (e-DMFT)
calculations~\cite{Georges-Kotliar-DMFTmethod-RMP1996,Kotliar-Haule-DMFTmethod-RMP2006,Haule-DMFTfullpot-PRB2010,Haule-DMFTdblecount-PRL15} 
 on LaNiO$_2$ and NdNiO$_2$. Our  calculations in the paramagnetic and magnetic states constitutes one of the very few fully-self-consistent e-DMFT study in these systems and reveal a rich variety of physics. The physical picture that emerges is depicted in the correlation (U) - temperature (T) phase diagram, Fig.~\ref{fig:UT_PD}.
Notable among several interesting features is a 
Fermi liquid (FL) phase at low temperature, a Curie-Weiss (CW) regime at high temperature, and an antiferromagnetic (AFM)  phase straddling the FL and CW phases over a region of U and T. 
While the FL phase onsets at some characteristic temperature, the temperature scale for full screening of the Ni-d moments is determined to be lower. In this contest, we explore the possibility of 
Nozieres Exhaustion Principle (NEP)~\cite{NEP-Annals1985,NEP2-EPJ1998} playing a role.
Overall, we study several key properties of LaNiO$_2$ and NdNiO$_2$. 

\section{Results}
\label{sec_Results}

\subsection{Phase diagram}

To construct the phase diagram, we perform calculations over a wide range of temperature (50K - 3500K) and correlation strength (U $\sim$ 5-15 eV). In panel Fig.~\ref{fig:UT_PD}(a), we show the overall U-T phase diagram and in panel (b), only the low temperature part. The phase diagram is constructed based on our calculated local magnetic susceptibility, that provides information about the existence of local moment on the Ni ions (see Fig. ~\ref{fig:UT_PD}(c)), and scattering rates, that give information about the electronic properties of a particular state - Fermi liquid versus non-Fermi liquid; see Fig.~\ref{fig:UT_PD}(d). As discussed later, these are supplemented by our study of 
density of states (DOS), orbitally projected spectral functions and hybridization, all as function of U and T.  

The phase boundary between the FL and CW phases is shown by the green triangles. To establish this boundary, we fit the local magnetic susceptibility computed for various values of the correlation parameter U (for fixed exchange interaction J = 1 eV) to the Curie-Weiss-Wilson expression shown in the caption to Fig.~\ref{fig:UT_PD}. In Fig.~\ref{fig:UT_PD}(c), we show a typical fit for the case 
of U = 7 eV. (Results and fits for other values of U are presented in Supplementary Information).
Broadly, there are three regimes: the low-T regime where the magnetic susceptibility deviates from the Curie-Weiss susceptibility behavior, and at a lower temperature, approaches a constant. We denote this as the Fermi liquid regime. 
We call the intermediate regime, where the magnetic susceptibility can be fitted with a Curie-Weiss-Wilson formula (caption, Fig.~\ref{fig:UT_PD}), the Curie-Weiss I (CWI) regime. We note that the screening temperature obtained from Wilson formula, T$^*$, is consistent with the temperature where the calculated magnetic susceptibility deviates from the CW behavior.  At very high T, the local magnetic susceptibility deviates from the typical Curie-Weiss form. This could be due to a change in the local moment (resulting in different slope of the CW law), or due to additional high-T corrections to the  Curie-Weiss-Wilson formula; this is labelled Curie-Weiss II (CWII) regime. We denote the temperature that separates CWI and CWII regimes as T$_{CW}$, shown as red squares. In Fig.~\ref{fig:UT_PD}(c), the Fermi liquid regime is represented by the black symbols, CWI by the green ones and CWII by the red symbols. 

To have a deeper understanding of the FL and CW phases, we studied the temperature dependent scattering rates for d orbitals of the Ni ions, see Fig.~\ref{fig:UT_PD}(d). As in the case of magnetic susceptibility, 
the temperature dependent scattering rate also shows three regimes:  A low-T regime where the scattering rate is very small (zero within the error bar of the calculation);  a second regime (still low-T) where the scattering rate has a T$^2$ behavior, a feature of a Fermi liquid; a third high-T regime where the scattering rate is large and tends towards a constant value. We call T$^{*’}$ the temperature below which the scattering rate has T$^2$ behavior and we see that this temperature corresponds to the temperature T$^*$  below which the local susceptibility deviates from CWI behavior. Thus the FL regime appears to be well established by both the scattering rate and the local magnetic susceptibility. We also observe that the temperature above T$^{*’}$, where the scattering rate tends to a constant value, is similar to T$_{CW}$. In addition, the temperature, T$_{FLFS}$, below which the scattering rate is practically zero, corresponds to the temperature scale for a constant local magnetic susceptibility. These results can be interpreted in the following way: at very high temperatures, there is a local moment at the Ni sites due to the d electrons. With lowering of the temperature, these local moments start to get partly screened around T$^*$, signaling the onset of FL behavior. The moments, however, get fully screened 
only below T$_{FLFS}$. This is denoted by blue stars in the phase diagram.

As can be seen in Fig.~\ref{fig:UT_PD}(a),(b) both the FL $\rightarrow$ CWI and the CWI $\rightarrow$ CWII crossover temperature scales decrease with increasing values of U, so that the FL regime is squeezed
down as U is increased. From the plots of $\chi(T)$ and scattering rate $\Sigma$ (Fig.~\ref{fig:UT_PD}(c),(d)), it can also be seen that the scattering rate tracks the behavior of susceptibility.
 We have checked that in the FL regime, the imaginary part of the self-energy, Im$\Sigma(i\omega_n=0) \rightarrow 0$ (as expected), while it is non-zero in the two CW regimes.
The temperature dependence of our spin susceptibility has some commonality with susceptibility experiments~\cite{Hayward99} that shows weak temperature dependence, and some evidence of flattening of a Curie-Weiss curve at lower temperatures. Also, the susceptibility was found to be significantly smaller than for a spin-1/2 paramagnet; the experimental data was shown not to fit to a S=1/2 paramagnet, and indicated that 
effective spin of the system may be smaller than 1/2. Our calculations indicate that  $S \le 1/2$ ($\sim$ 0.2-0.3) at low temperatures. 

\begin{figure*}[!htb]
\includegraphics[width=0.98\linewidth]{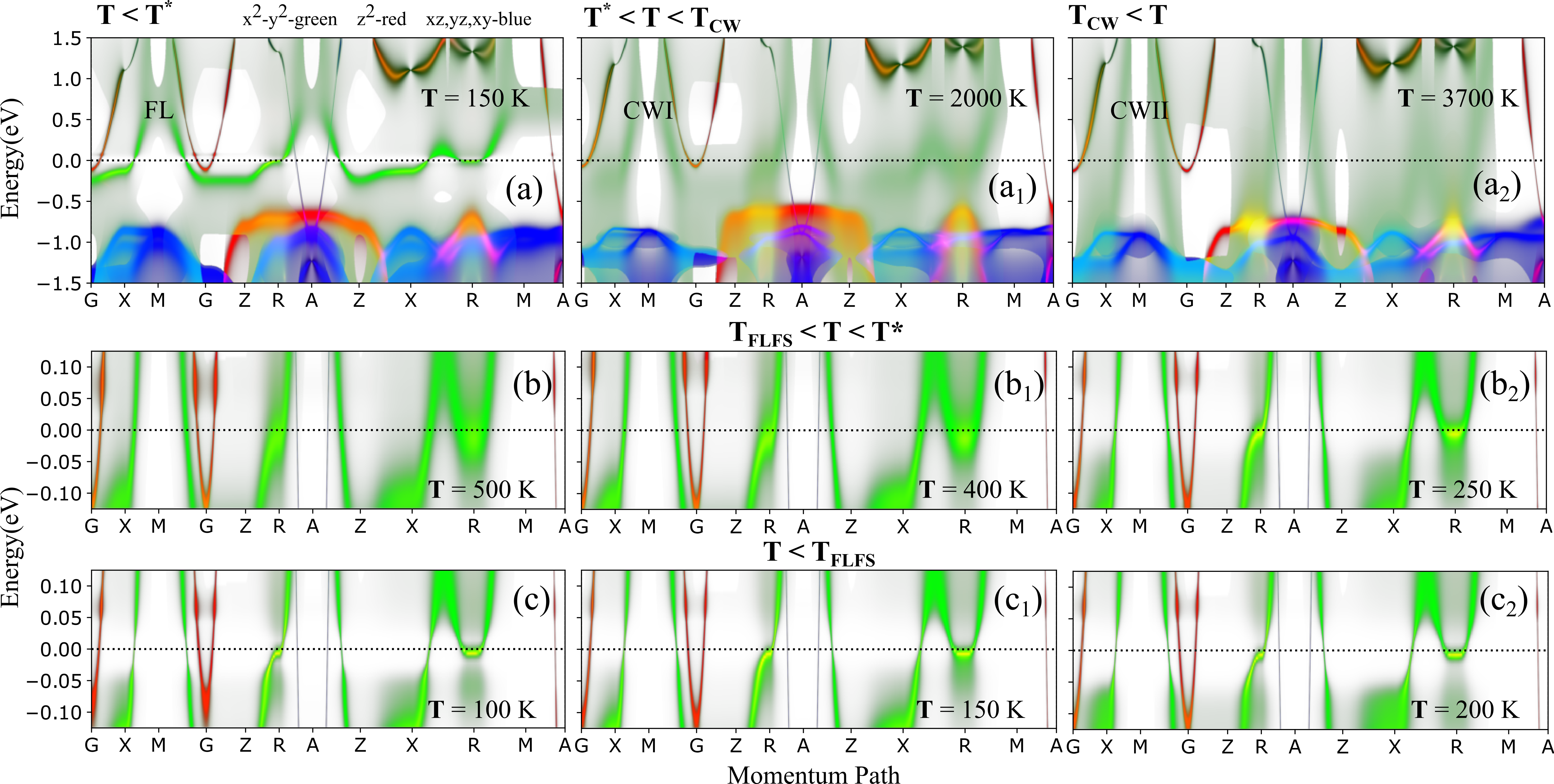}
\caption{(Color online)  \textbf{Orbital projected Spectral functions for LaNiO$_2$;  U = 7 eV :} Panels (a), (a$_1$) and (a$_2$) shows the temperature (T) dependence of the spectral function while the other panels show the temperature (T) dependence of the spectral function around the Fermi level for T$_{FLFS}$ $<$ T $<$ T$^*$ (panels (b), (b$_1$), (b$_2$)), and for T $<$ T$_{FLFS}$ (panels (c),( c$_1$), (c$_2$)). In the last set of panels, we can see explicitly the high degree of coherence of the states at the Fermi level.}
\label{fig:SWQLa}
\end{figure*}

\begin{figure*}[!htb]
\includegraphics[width=0.98\linewidth]{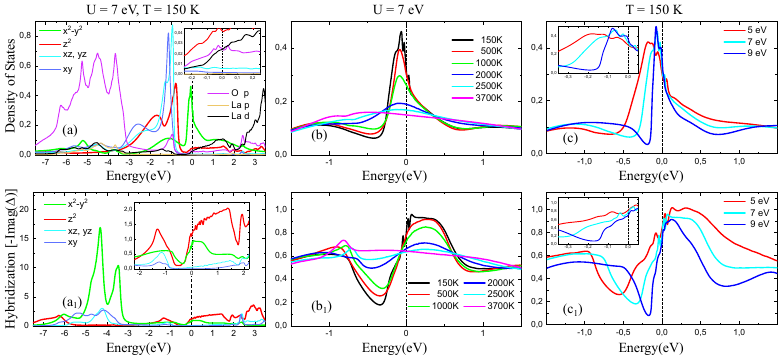}
\caption{(Color online)  \textbf{Density of states (DOS) for LaNiO$_2$:} Panel (a) shows the projected DOS for the Ni-d orbitals, O-p, La-p and La-d, while panel (a$_1$) shows the hybridization corresponding to the Ni-d orbitals.  Panels (b) and (b$_1$) shows the temperature (T) dependence of the Ni-$d_{x^2-y^2}$ projected DOS and the corresponding hybridization, while  panels (c) and (c$_1$) shows the correlation (U) dependence of the same Ni-d projected DOS and the corresponding hybridization. The insets focus on the energy region around the Fermi level, which is marked by the zero on the energy axis.}
\label{fig:DOSLa}
\end{figure*}

\begin{figure*}[!htb]
\includegraphics[width=0.98\linewidth]{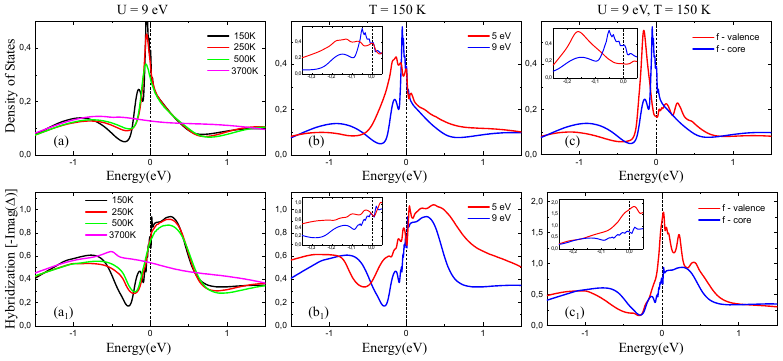}
\caption{(Color online)  \textbf{Density of states (DOS) for NdNiO$_2$:} Panels (a) and (a$_1$) show the temperature (T) dependence of the Ni-$d-{x^2-y^2}$ projected DOS and the corresponding hybridization, while panels (b) and (b$_1$) show the correlation (U) dependence of the Ni-$d_{x^2-y^2}$ projected DOS and the corresponding hybridization. Panels (c) and (c$_1$) shows the "effects of doping" on the $d_{x^2-y^2}$ DOS and its corresponding hybridization, by placing the Nd-f electrons into valence or into the core. The insets focus on the energy region around the Fermi level, which is marked by the zero on the energy axis.}
\label{DOS_Nd}
\end{figure*}

\begin{figure*}[!htb]
\includegraphics[width=0.98\linewidth]{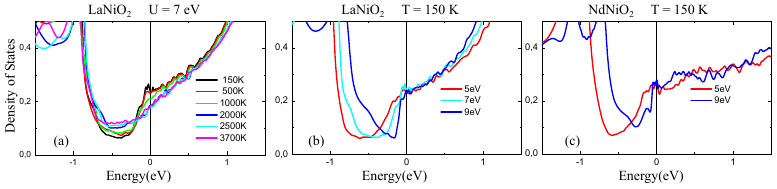}
\caption{(Color online)  \textbf{Density of states (DOS) per unit cell for all non-correlated (bath) electrons:} Panels (a) and (b) show the temperature (T) and correlation (U) dependence respectively for LaNiO$_2$, while panel (c) shows the correlation (U) dependence of DOS for NdNiO$_2$. }
\label{DOSsp}
\end{figure*}

\subsection{Spectral Function, Density of States, Hybridization}

To understand the electronic properties from a different perspective, we also computed the orbitally projected spectral function at various temperatures. Fig.~\ref{fig:SWQLa}(a) shows the results for U=7 eV.
There is a clear consistency between the scattering rates and the spectral functions for the three regimes, FL, CWI and CWII. For example, the Ni-d states are coherent and situated around the Fermi level in the FL phase while they are very incoherent in the CWI and CWII phases as expected based on the scattering rates. At the higher temperatures (CWII regime), the spectral function is almost washed out, implying
that there are no quasiparticles, just fluctuating local moments.
In addition, if we zoom in around the Fermi level, we observe that the Ni-$d_{x^2-y^2}$ band progressively sharpen with lowering of T,  and below T$_{FLFS}$, becomes very sharp, i.e.  highly coherent as expected due to the vanishing scattering rate; see Figs.~\ref{fig:SWQLa}(b)-(b2) and (c)-(c2). It may be noted that the Ni-$d_z^2$ spectral function does not change much at all across the entire temperature range.

Since the spectral function shows the electronic states just along several paths in the Brillouin zone, we also plot in Fig.~\ref{fig:DOSLa}(a) the partial density of states (pDOS), for U=7eV at T=150K (plots for other U's are presented in Supplementary Information), for all the Ni-d, O-p and La-p and La-d orbitals, together with the hybridization corresponding to the Ni-d orbitals in Fig.~\ref{fig:DOSLa}(a1). The pDOS for the Ni-d orbitals show that the center of mass for the pDOS for Ni $d_{x^2-y^2}$ orbital is around the Fermi level, while for the other d orbitals, the center of mass is at lower energies; thus these orbital make only a small contribution at the Fermi level. 
We find that the Ni-d states hybridize more with the O-p states and less with La-p or La-d states (reasoning based on results shown in Supplementary Information). 

Next, we consider the temperature dependence of the total and partial density of states. In Fig.~\ref{fig:DOSLa}(b) and (b$_1$), we show the temperature dependence for the Ni-$d_{x^2-y^2}$ DOS, and its corresponding hybridization around the Fermi level. The other orbitals do not show a strong temperature dependence around the Fermi level (see Supplementary Information). We observe that at low T, the $d_{x^2-y^2}$ pDOS has a peak structure just below the Fermi level; the peak structure decreases with increasing temperature, and eventually becomes rather flat. Similarly, at low T, the corresponding hybridization has 
dip below Fermi level accompanied by a peak above; this feature become less pronounced  with increasing temperature. The same type of behavior is seen at low temperature (shown for T=150K) for other values of U; see Fig.~\ref{fig:DOSLa}(c) and (c$_1$). With increasing U, the $d_{x^2 - y^2}$ DOS peak around $E_F$ sharpens, and the hybridization
with other orbitals decreases to some extent. 

These feature of the $d_{x^2-y^2}$ pDOS and hybridization bear resemblance to results from calculations on a periodic Anderson model with an f and s electrons~\cite{Meyer-Nolting-PRB2000}. In that case, at T=0, the 
system shows a gap in the DOS at half filling for both s and f bands (with peaks in the DOS around the gap), accompanied by peak in hybridization within the gap.
The gap in the DOS and the peak in the hybridization progressively disappear upon reducing the number of s-electrons. The f-electron pDOS shows an asymmetric peak shape around the Fermi level while the hybridization exhibits a dip followed by a peak, similar to an asymmetric sinusoidal curve (see Figs. 5 and 6 in Ref. [\onlinecite{Meyer-Nolting-PRB2000}]).
In addition, there are similarities between the high-T model calculations of aforementioned work and our eDMFT calculations; in both cases the impurity electron pDOS is flat around the Fermi energy.
Reducing the number of s electrons results in an insufficient number of electrons to screen the f electrons; this was associated with Nozieres Exhaustion Principle~\cite{NEP-Annals1985,NEP2-EPJ1998}. Comparing the U dependence of our DOS and hybridization for the $d_{x^2-y^2}$ orbital (Fig.~\ref{fig:DOSLa}(c) and (c$_1$)), with that of Ref. [\onlinecite{Meyer-Nolting-PRB2000}] as function of decreasing s-electrons, we see a similar trend. This can be interpreted based on the local moment shown in the inset of our Fig.~\ref{fig:UT_PD}(a). Since the total number of electrons in the system is conserved, increase of the local magnetic moment with increasing U means that the number of bath electrons that are left to screen the moment is decreasing. 

To further explore screening of Ni-d electrons, besides LaNiO$_2$, we also looked at the same properties for NdNiO$_2$, see Fig.~\ref{DOS_Nd}. The difference between LaNiO$_2$ and NdNiO$_2$ is that Nd has 3 f-electrons. If we place the f electrons into the core in our calculations, we see that the temperature dependence of the DOS and hybridization are very similar to that of LaNiO$_2$; see Fig.~\ref{DOS_Nd} (a) and (b). Since our results for LaNiO$_2$, in some ways, resemble that of the periodic Anderson model (where NEP was suggested), we explored this aspect further in the case of NdNiO$_2$. We can consider 
the f electrons as valence ones or in the core, as mimicking scenarios where we dope the system with holes or electrons. For example, if we start with the f electrons as valence electrons, when we put them into core, we can think of this system as a model system which is doped with holes. This is exactly the scenario described in Ref. [\onlinecite{Meyer-Nolting-PRB2000}], when the number of s electrons is reduced while the number of f electrons in kept fixed. The DOS and hybridization in  Fig.~\ref{DOS_Nd} (c) and (c1) have the same trends as the corresponding quantities in the periodic Anderson model calculation~\cite{Meyer-Nolting-PRB2000}. In addition, we also looked at the T and U dependence of the non-correlated electrons and they have the same trends as the s electrons in the above-mentioned work (see our Fig.~\ref{DOSsp}). 
We also note that the Ni-d electron hybridization (Fig.~\ref{fig:DOSLa}(b1) and the bath electrons DOS (Fig.~\ref{DOSsp})
exhibit progressively larger peak with decreasing temperature, pointing to better screening at the lower temperatures.

\subsection{Nickel-d Occupancy}

\begin{figure}[!htb]
\includegraphics[width=0.95\linewidth]{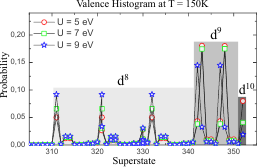}
\caption{(Color online) \textbf{Ni-d electron Valence Histogram}}
\label{fig:histogram}
\end{figure}

To understand the increase of the local effective moment in the CW regime with increasing U, we also looked the Ni-d valence histogram, see Fig.~\ref{fig:histogram}. By studying the valence histogram we observe that the system is in a mixed d$^8$-d$^9$ configuration, with the d$^9$ configuration being favored for small U and d$^8$ for large U values. 

There has been ongoing discussions/debate on the valence character of the Ni-d electron, in particular whether the character is more like Ni-$d^8$ or Ni-$d^9$. Based on our e-DMFT calculations,
 the occupancy of the Ni-d electrons, $n_d(U)$ changes from $\sim$8.6e  to $\sim$8.35e as we go from U=5 eV to U= 9 eV. 
The occupancies can be better understood on considering Fig. ~\ref{fig:histogram}, which shows the probability of DMFT sampling of Ni-d superstates (configurations of microstates) for U=5, 7 and 9. 
Overall, we find  higher probabilities for $d^9$ vs $d^8$ configurations for U=5 vs U=9, thereby giving $n_d \sim$8.6 (closer to 9) for U=5 eV, and $n_d \sim$ 8.35 (closer to 8) for U=9 eV.

\subsection{Magnetism}

In addition to our paramagnetic (PM) calculations discussed above, 
we also carried out DMFT calculations in the ferromagnetic (FM) and in-plane checkerboard antiferromagnetic (AFM) configurations  for several values of U between 5 and 15 eV.
The AFM structure is not stable for $U \le 7$ eV. For U $>$ 7 eV, and up to 12 eV, it is stable in a certain range of temperatures shown in Fig.~\ref{fig:UT_PD}(a),(b) (black dashed line) with ordered moment corresponding to $|S_z| \sim$ 0.3, demonstrating a dome-like feature. (See Supplementary Information for results supporting this). We note that the AFM state is only slightly lower in energy than the PM state, the energy difference 
 $\sim$ 5 meV, being within the DMFT error bar. 
The FM state is not stable for U = 5-12 eV, but may be stable for U $\sim$13-15 eV. 
Our result, for $U \le 7$ eV, is consistent with neutron scattering experiment~\cite{Hayward99,Hayward03}, and $\mu$sr/magnetic susceptibility measurements~\cite{Ortiz-PRR22} 
that also do not find long-range magnetic order.
We note that even with increasing values of U, in contradistinction to some of other work~\cite{Gu-comphys20,Karp-PRX20},
we do not find any insulating phase. However, it is possible that the Ni-$d_z^2$ orbital becomes progressively more metallic with increasing U, while $d_{x^2-y^2}$ gets less metallic. This may indicate some form of orbital selectivity that has been suggested~\cite{Lechermann-PRB20}. 

\subsection{Effective Mass and Specific Heat}

Within DMFT, the quasiparticle renormalization factor, $Z$ for the impurity atom orbitals can be obtained from the derivative of the imaginary part of the self-energy at the Fermi energy ($\omega_n = 0$). The corresponding effective mass  $m^*/m_{DFT} = 1/Z$, $Z$ being the quasiparticle renormalization. At T = 150 K (in the regime we have characterized as Fermi liquid),
we find that for U = 5, 7, 9 eV, $m^*/m_{DFT}$  for the $d_{x^2 - y^2}$ orbital is 2.8, 4.3, 6.5 respectively, and for the $d_z^2$ orbital 1.3, 1.5, 1.7 respectively. This, together with our calculated DOS at $E_F$, can be used to estimate the Ni-$d$ electron contribution to the linear coefficient of the specific heat, given by: $\gamma = const \sum_{orbital} N(0)_{\rm orbital}/Z_{\rm orbital}$  mJ mol$^{-1}$ K$^{-2}$; where N(0)$_{\rm orbital}$ is the density of states of an orbital at the Fermi energy, with units of 1/(eV/atom); $Z_{orb}$ is the renormalization factor of orbital being summed over; and the $const$ = 2.349  results from unit conversion. The $t2g$ orbitals are all occupied states and have no weight at the Fermi level. Hence, these do not enter into our estimate of 
$\gamma$. We estimate $\gamma \sim$ 2.5, 3, 6 mJ mol$^{-1}$ K$^{-2}$ for U = 5, 7, 9 eV respectively. Recent specific measurements~\cite{Ortiz-PRR22} on LaNiO$_2$ find a linear-T behavior with a possible $T^3 \ln T$ correction, reflecting Fermi liquid behavior at low temperature.  The experimental  $\gamma \sim$ 4-5 mJ mol$^{-1}$ K$^{-2}$ are consistent with our estimates of $\gamma$. We note that our estimates are based solely on the contribution of the Ni-$d$ orbitals and do not take into account O or La orbitals. Since the DOS of O-p and La-d orbitals at $E_F$ are very small, these would give negligible additional contribution ($\sim$ 0.1) to the total $\gamma$.

\section{Discussion}
\label{sec_Discussion}

Our first principle DFT + DMFT (e-DMFT) calculations are fully self-consistent in that the charge is recalculated after a DMFT run and then iterated through the DFT and DMFT loops till 
convergence is achieved. This is one of the very few fully self-consistent calculations performed on the nickelate systems, and thus differs from other 
flavors of DMFT that have been used, viz., Wannier down-folded DMFT,  DMFT on models, and single-shot DMFT (hence not fully self-consistent). We do not employ downfolding and use a
a large hybridization window of $\pm$ 10 eV around the chemical potential. Another aspect of our calculations is the use of 'exact double-counting'~\cite{Haule-DMFTdblecount-PRL15}, which
has been known to increase the precision of the calculations.

There are important differences between our findings and those of some of the other DMFT work~\cite{Gu-comphys20,Karp-PRX20} on LaNiO$_2$ and NdNiO$_2$. We do not find a stable AFM phase for
 $U \le 7$ eV. An AFM phase is found to be barely stable for U $>$ 7 (up to U=11 eV), but with energy difference from the PM phase lying within the error bar of the DMFT calculations. 
 For NdNiO$_2$, Ref. [\onlinecite{Karp-PRX20}] finds stable AFM phase phase between U $\sim$ 2.3 eV to U $\sim$ 8 eV, with a crossover from AFM metal to AFM insulator between U $\sim 5.7$ and 8 eV.
 Also, in contradistinction with others, we do not find any insulating phase, even for large values of U up to 15 eV. Thus, we are at odds with Ref. [\onlinecite{Gu-comphys20}] that find an insulating phase for U = 7 eV, 
and Ref. [\onlinecite{Karp-PRX20}]  that obtains an insulating PM behavior in NdNiO$_2$ for $U > 7 eV$. We note that the above are down-folded  DMFT calculations, and 
 single-shot (so self-consistency not achieved between DFT and DMFT loops), and that could be possible reason for results that are different from ours. Additionally, some of the calculations
 are on the Nd compound, though it has been pointed out that the Nd-f electrons may not play a crucial role in the nickelate properties; Ref. [\onlinecite{Gu-comphys20}] does not find much difference between 
 the La and the Nd compound.
 
DFT and DFT+U studies~\cite{Savrasov-DMFT-PRB20,Moritz-DFT-PRX21} find a second band, comprising of rare-earth $5d$ orbitals crossing the Fermi energy; a charge transfer energy appreciably larger than in the cuprates, and a antiferromagnetic state for certain values of the correlation $U$. While some of these aspects are also found in down-folded DMFT calculations~\cite{Karp-PRX20,Gu-comphys20,Lechermann-PRB20}, there are differences between these and the DFT results. One variance pertain to the nature of a second band crossing $\epsilon_F$. Some of the above-mentioned down-folded DMFT work, such as Ref. [\onlinecite{Karp-PRX20}], a  DFT + sicDMFT calculation~\cite{Lechermann-PRB20}, and a self-consistent DMFT calculation~\cite{Wang-Kotliar-PRB20}  finds that that the second band has mixed rare-earth (La or Nd)-$d_z^2$
and Ni-$d_z^2$ character at the $\Gamma$ point and rare-earth Nd-$d_{xz}$ and Ni-$d_{yz}$ character at the $A$ point; these form small electron pockets. These are at odds with 
Ref. [\onlinecite{Gu-comphys20}] that finds that the second band does not arise from hybridization with the rare-earth orbital, rather with interstitial $s$- electrons, mediated by O-$p$ orbitals. 
DFT and some DMFT calculations~\cite{Lechermann-PRB20,Wang-Kotliar-PRB20} find the Ni-$3d$-O-$2p$ hybridization to be small, and others, such as, Ref. [\onlinecite{Gu-comphys20}] find this to be strong, 
though in Ref. [\onlinecite{Wang-Kotliar-PRB20}], this varies depending on the rare-earth in question. The down-folded versions of DMFT typically find insulating behavior for sufficiently large U (in some temperature range). Ref. [\onlinecite{Lechermann-PRB20}] suggests a Mott gap for the $d_{x^2 - y^2}$ orbital for  $U \ge 10$ eV.

We do not find evidence for the Hund's metal scenario~\cite{Kang-Kotliar-Hunds-PRL21} that has been proposed based on study of optical conductivity at intermediate energies. In our study at low energies, for a fixed U, we find weak dependence of the spectral function and spin susceptibility as the exchange interaction parameter is varied from 1 eV to 0.1 eV; the variation with U is much larger (see Supplementary Information for details). 

The picture of the parent nickelate compound being a Nozieres type Fermi liquid at low temperature emerges from our detailed calculations and careful analyses of magnetic susceptibility, quasiparticle scattering rate,
and spectral function. The progression of the Fermi liquid phase from one in which the Ni-moments are partially screened to one wherein the moments are fully screened can be attributed to features in the temperature-dependent hybridization of the correlated Ni-$d_{x^2-y^2}$ electrons with the non-correlated electrons that include the rare-earth-d, O-p and other conduction electrons that comprise the "bath". We speculate that this
picture of the screened FL may have consequences for superconductivity in the doped compound.

 \section{Conclusion}
 
 In summary, we have carried out a fully self-consistent first principle embedded DMFT calculations in the paramagnetic and magnetic states of infinite layer parent nickelate compounds $R$NiO$_2$, for $R$ = La and Nd. Our results and conclusions are based on detailed calculations  of the electronic structure, local spin susceptibility, and scattering rate for a sizable set of correlation strength, and wide range of temperature. In our picture, the nature and temperature dependence of hybridization, together with the temperature dependence of density of states are important for understanding the physics in different temperature regimes. 
The array of interesting many-body ground states and correlation and temperature scales that emerge are depicted in our proposed correlation-temperature phase diagram.
The overall broad features are a Fermi liquid phase at low temperature, a Curie-Weiss regime at high temperature, and an antiferromagnetic phase  over a narrow region of U and T. 
The onset of the Fermi liquid regime at a characteristic temperature $T^*$  is signaled by $T^2$ quasiparticle scattering rate and deviation from Curie-Weiss behavior in local spin susceptibility.
Based on scattering rates, local spin susceptibility, and orbital projected spectral function, we believe this to occur when the Ni-d moments start to undergo Kondo screening resulting in a Nozieres type Fermi liquid. 
A Kondo screening picture of the ground state of the parent infinite layer nickelate compound has been suggested, based on X-ray spectroscopy experiments.~\cite{Hepting-Natmat20}  Interestingly, upon closer analysis, some finer features are revealed. Thus, the Ni-d moments get fully screened only at a lower temperature, $T_{FLFS}$, correlating with a constant spin susceptibility, a scattering rate 
that goes to zero, and an extremely sharp $d_{x^2 - y^2}$ band feature near the Fermi level. This picture of screening is consistent with the Ni-d electron hybridization and the bath electrons DOS
exhibiting progressively larger peak with decreasing temperature. Also, there is deviation from the typical Curie-Weiss behavior at relatively high temperature, the scale for 
which is $T_{CW}$, and is signaled by the scattering rate flattening towards a constant, and deviation of spin susceptibility from the CW behavior.

Although the system is already exhibiting Fermi liquid features starting at the higher temperature, $T^*$, the lower temperature scale for full screening of the Ni-d moments may be pointing to insufficiency of the bath electrons to effectively screen the Ni-d moments till a lower temperature is reached, resulting
in what we have called fully screened Fermi liquid (FLFS). 
This may be suggestive of the Nozieres Exhaustion Principle~\cite{NEP-Annals1985,NEP2-EPJ1998} at play, as suggested in DMFT calculations in metallic Nickel~\cite{Nickel-Natcom17},
and in periodic Anderson model calculations~\cite{Meyer-Nolting-PRB2000}  with an f and s electrons. 

Our finding of a Fermi liquid phase at low temperature is consistent with specific heat measurement on LaNiO$_2$ showing a 
linear-T behavior,  On the question of single-band versus multi-band governing the physics of the nickelates,
our work finds that in addition to the dominant $d_{x^2-y^2}$ band, the presence of a second band comprising of hybridized $d_z^2$ band around the $\Gamma$ point. The second band contributes
about 1/5 of the total Ni-d moment at low temperature (at 200 K). However, as our calculated temperature-dependent DOS, hybridization and spectral function show, the $d_z^2$ band does not show 
much temperature dependence. Thus, in our proposed picture, at higher temperatures, the physics would be governed by the single $d_{x^2-y^2}$ band.
Based on the lack of experimental evidence for long-range magnetic order down to low temperatures, and the observation of 
magnetic excitations in NdNiO$_2$~\cite{Lu-magexcit-Science22}, our calculations would place the $R$NiO$_2$ materials close to the dome of anti-ferromagnetism in our phase diagram, thereby making antiferromagnetic fluctuations feasible. 
In particular, our study indicate that for LaNiO$_2$, a non-magnetic state (as indicated in Refs. [\onlinecite{Hayward99,Hayward03,Ortiz-PRR22}]),
with $n_d$ closer to $d^9$ ($\sim$ 8.6) can be achieved with U = 5-7 eV.
Separating non-magnetic from magnetic behavior (see under Magnetism in Supplementary Information), U=7eV  appears to be a "special point " in the phase diagram
for LaNiO$_2$. The same may be true for the Nd- or Pr-nickelate, with a different value of U as the characteristic scale for the onset of AFM. Our constrained DMFT calculations gives U= 6.5 eV for LaNiO$_2$. We do not find evidence for insulating behavior up to U=15 eV.
 
\section{Methods}
\label{Sec_Methods}

The flavor of DMFT we use is embedded dynamical mean-field theory (e-DMFT)~\cite{Georges-Kotliar-DMFTmethod-RMP1996, Kotliar-Haule-DMFTmethod-RMP2006,Haule-DMFTfullpot-PRB2010,Haule-DMFTdblecount-PRL15}, developed at Rutgers. We do not employ downfolding, and it is thus distinct from Wannierized or downfolded DMFT versions. For the DFT part, i.e. obtaining the associated electronic band structure, we performed the full-potential linearized augmented plane wave (FP-LAPW) method as implemented in the WIEN2k code~\cite{WIEN2K-Blaha2020}. Our calculations are 'fully self-consistent' between the DFT and DMFT parts, and employs 'exact double-counting'~\cite{Haule-DMFTdblecount-PRL15}. We use the  'Full Coulomb' option (as opposed to 'Ising'), i.e. the spin-rotation invariant form of the Coulomb repulsion of the Slater 
form.~\cite{Haule-DMFTfullpot-PRB2010} We use a large hybridization window of $\pm$ 10 eV. For the DMFT projectors, we choose quasi-atomic localized orbitals. The radial part is the solution of the Schroedinger equation in the muffin-tin sphere, with linearized energy at the Fermi level, and the angular dependence is given by the spherical harmonics. For the quantum impurity problem (Ni-d electron), we use a version of continuous time quantum Monte Carlo (CTQMC) impurity solver. The needed analytic continuation from imaginary to real frequency axis is done using the maximum entropy method. We take the well-known tetragonal crystal structure for RNiO$_2$  with the rare earth element located at the center of the unit cell (see Fig in Supplementary Information); the related space group is P4/{\it mmm}.

\section{Acknowledgements}
\label{sec_Acknowledgements}

G.L.P.'s and L.C.'s  work were supported by a grant of the Romanian Ministry of Education and Research, CNCS - UEFISCDI, project number PN-III-P1-1.1-TE-2019-1767, within PNCDI III. K.H. were supported by the  U.S. Department of Energy, Office of Science, Basic Energy Sciences, as a part of the Computational Materials Science Program, funded by the  U.S. Department of Energy, Office of Science, Basic Energy Sciences, Materials Sciences and Engineering Division. KQ  acknowledges a QuantEmX grant from ICAM and the Gordon and Betty Moore Foundation, Grant GBMF5305, which partly funded this work.

\bibliography{Bib_Nickelates.bib}

\end{document}


\pagestyle{empty}

\title{ Supplementary Information for "Correlation-temperature phase diagram of prototypical infinite layer rare earth nickelates"}
\maketitle



\section{Supplementary 1: Correlation (U) dependent local magnetic susceptibility}

\begin{figure*}[!htb]
\includegraphics[width=0.98\linewidth]{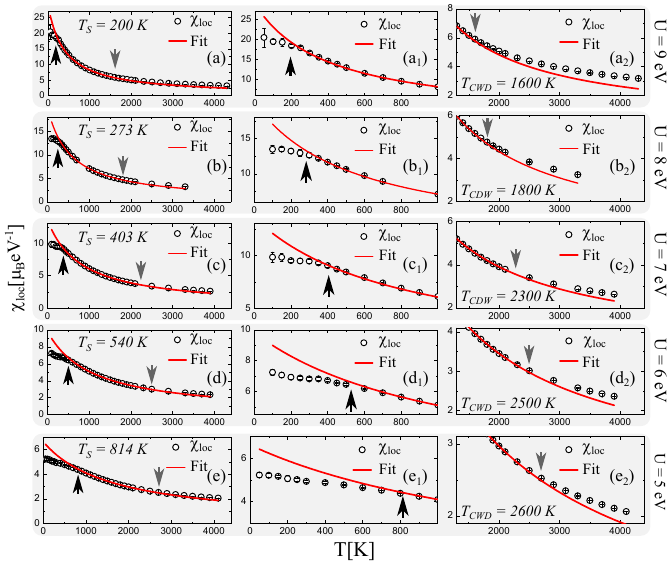}
\caption{(Color online)  \textbf{Correlation (U) dependent local magnetic susceptibility for LaNiO$_2$:} Panels (a) to (e) shows the temperature dependence of the theoretical local magnetic susceptibility ($\chi_{loc}$) for the d electrons of the Ni ions for U = 5-9 eV and J = 1 eV. Panels (a$_1$) to (e$_1$) and (a$_2$) to (e$_2$) shows $\chi_{loc}$ with the focus on the low- and high- temperature range. The solid line represents the best fit with the Curie-Weiss-Wilson formula $\chi_{loc} = \frac{C \mu^2_{eff}}{(T + 2 T_S)}$. From panels (a$_1$) to (e$_1$) and (a$_2$) to (e$_2$) we observe that $T_S$ and $T_{CDW}$ corresponds to the temperatures where the $\chi_{loc}$ shows deviations from the  Curie-Weiss behavior.  The arrows show the position of various critical temperatures.}
\label{Sus_U_T_fits}
\end{figure*}

\newpage

\section{Supplementary 2: Correlation (U) dependent scattering rate }

\begin{figure*}[!htb]
\includegraphics[width=0.98\linewidth]{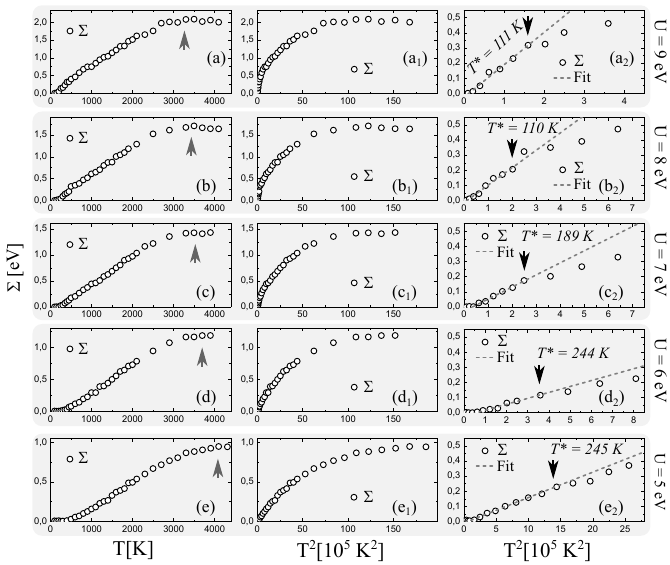}
\caption{(Color online)  \textbf{Correlation (U) dependent  scattering rate for the Ni $d_{x^2-y^2}$ orbital in LaNiO$_2$:} Panel (a) to (e) shows the temperature dependence of the theoretical scattering rate ($\Sigma$) for the electrons in the $d_{x^2-y^2}$ orbital of the Ni ions for U =  5 - 9 eV and J = 1 eV. Panels (a$_1$) to (e$_1$) and (a$_2$) to (e$_2$) focus on the low-temperature, where $\Sigma$  is plotted versus T and $T^2$. The dash line represents the best linear fit with the formula $\Sigma = A T^2 - B$, where A, B $>$ 0. $T^{*} = \sqrt{\frac{B}{A}}$ is the temperature at which there is onset of Fermi liquid behavior (linear dependence of $\Sigma$ on T$^2$).  The arrows show the position of various critical temperatures.}
\label{Scatt_U_T_fits}
\end{figure*}

\newpage

\section{Supplementary 3: Correlation (U) and temperature (T) dependence of  local magnetic susceptibility and scattering rate}

\begin{figure*}[!htb]
 \begin{adjustbox}{addcode={\begin{minipage}{\width}}{ \caption{(Color online)   \textbf{Correlation and Temperature dependent  local magnetic susceptibility and scattering rate for LaNiO$_2$:} These plots are a repetition of some of the plots from Figs. \ref{Sus_U_T_fits} and \ref{Scatt_U_T_fits} aiming to give a better comparison of various temperature scales obtained from the local magnetic susceptibility and the scattering rate.} \end{minipage}},rotate=270,center}
\includegraphics[scale=2.0]{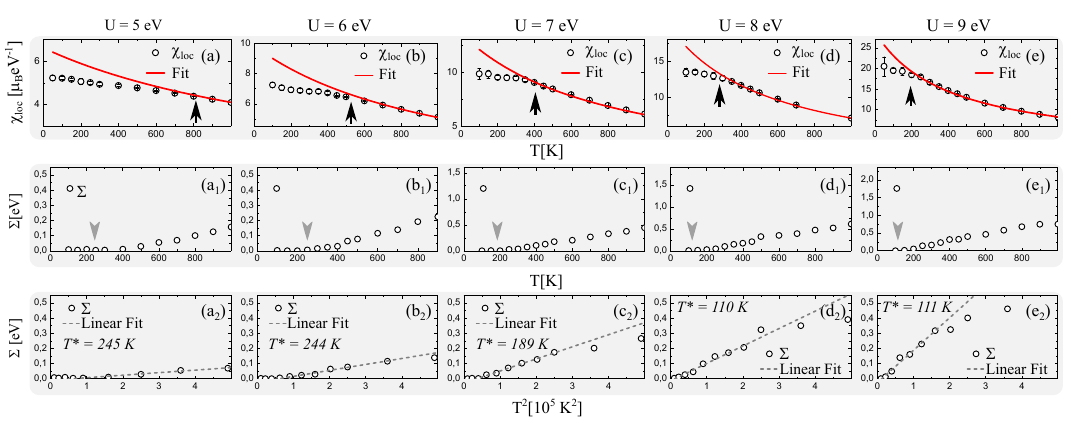}
\label{Chi_Scatt_U_T_fits}
  \end{adjustbox}
\end{figure*}

\section{Supplementary 4: Correlation dependent Density of states (DOS) for ${\bf LaNiO_2}$}

\begin{figure*}[!htb]
\includegraphics[width=0.98\linewidth]{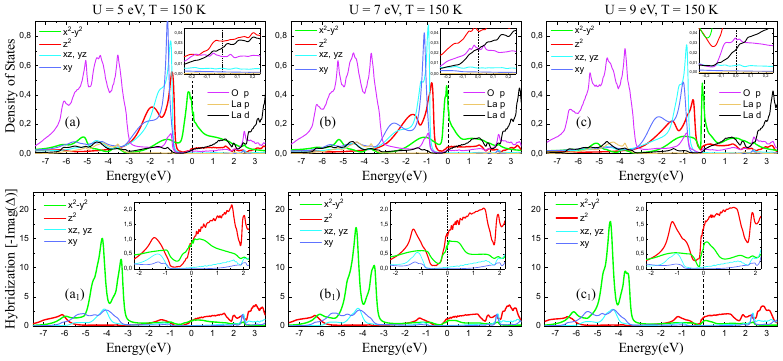}
\caption{(Color online)  \textbf{Correlation dependent Density of states (DOS) for LaNiO$_2$:} Panel (a), (b) and (c) shows the projected DOS for the Ni-d orbitals, O-p, La-p and La-d while panels (a$_1$), (b$_1$) and (c$_1$),  shows the hybridization corresponding to the Ni-d orbitals. The insets focus on the energy region around the Fermi level marked by the zero on the energy axis.}
\label{DOS_La_z2}
\end{figure*}

\section{Supplementary 5: Correlation and Temperature dependent Density of states (DOS) and hybridization for ${\bf LaNiO_2}$}

\begin{figure*}[!htb]
\includegraphics[width=0.98\linewidth]{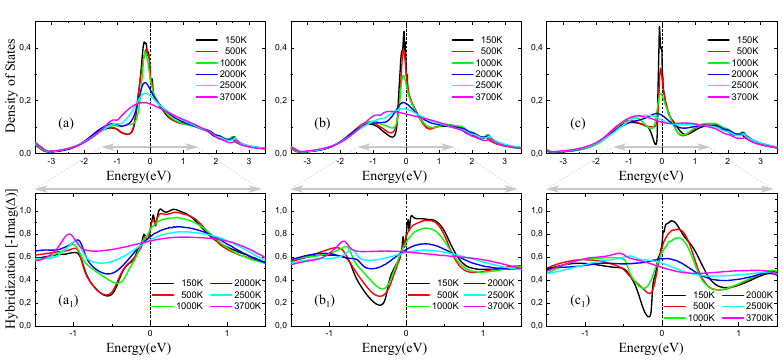}
\caption{(Color online)  \textbf{Correlation and Temperature dependent Density of states (DOS) for LaNiO$_2$:} Panels (a), (b) and (c) shows the temperature (T) dependence of the $d_{x^2-y^2}$ projected DOS and the corresponding hybridization in panels (a$_1$), (b$_1$) and (c$_1$), for U = 5, 7 and 9 eV.}
\label{DOS_La_z2}
\end{figure*}

\section{Supplementary 6: Correlation and Temperature dependent Density of states (DOS) and hybridization for ${\bf NdNiO_2}$ with $f$-electrons in core}

\begin{figure*}[!htb]
\includegraphics[width=0.98\linewidth]{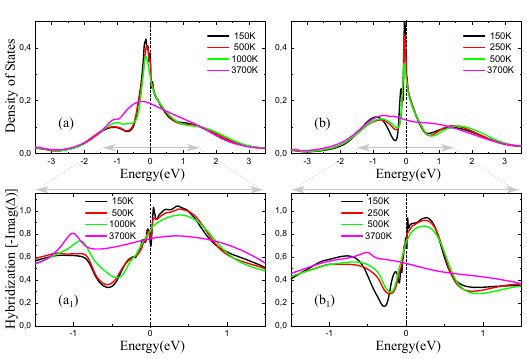}
\caption{(Color online)   \textbf{Correlation and Temperature dependent Density of states (DOS) for NdNiO$_2$ with Nd-f electrons placed into core:} Panels (a) and (b) shows the temperature (T) dependence of the 
$d_{x^2-y^2}$ projected DOS and the corresponding hybridization in panels (a$_1$) and (b$_1$), for U = 5 and 9 eV, J = 1 eV.}
\label{DOS_La_z2}
\end{figure*}

\newpage

\section{Supplementary 7: Correlation and Temperature dependent Density of states (DOS) and hybridization for ${\bf NdNiO_2}$ with $f$-electrons in valence}

\begin{figure*}[!htb]
\includegraphics[width=0.98\linewidth]{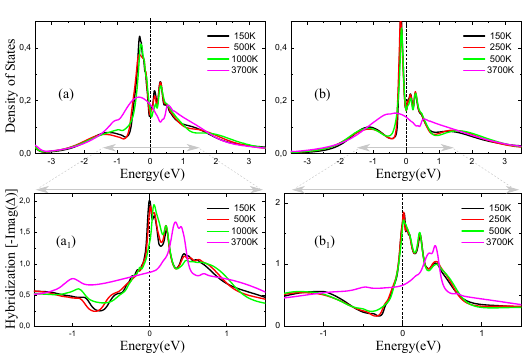}
\caption{(Color online)  \textbf{Correlation and Temperature dependent Density of states (DOS) for NdNiO$_2$ with Nd-f electrons placed into valence:} Panels (a), (b) and (c) shows the temperature (T) dependence of the $d_{x^2-y^2}$ projected DOS and the corresponding hybridization in panels (a$_1$), (b$_1$) and (c$_1$), for U = 5, 7 and 9 eV.}
\label{DOS_La_z2}
\end{figure*}

\newpage

\section{Supplementary 8: Correlation and Temperature dependent Density of states (DOS) and hybridization for $d_z^2$ orbital in  ${\bf LaNiO_2}$}

\begin{figure*}[!htb]
\includegraphics[width=0.98\linewidth]{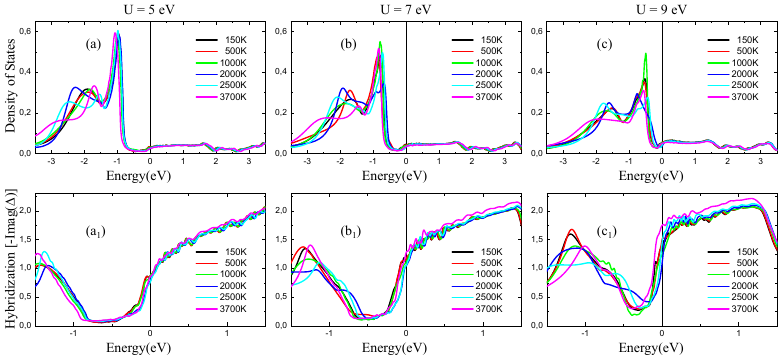}
\caption{(Color online)  \textbf{Correlation and Temperature dependent Density of states (DOS) for LaNiO$_2$:} Panels (a), (b) and (c) shows the temperature (T) dependence of the $d_z^2$ projected DOS and the corresponding hybridization in panels (a$_1$), (b$_1$) and (c$_1$), for U = 5, 7 and 9 eV.}
\label{DOS_La_z2}
\end{figure*}

\newpage

\section{Supplementary 9: On Hund's Metal Scenario}

\begin{figure*}[!htb]
\includegraphics[width=0.98\linewidth]{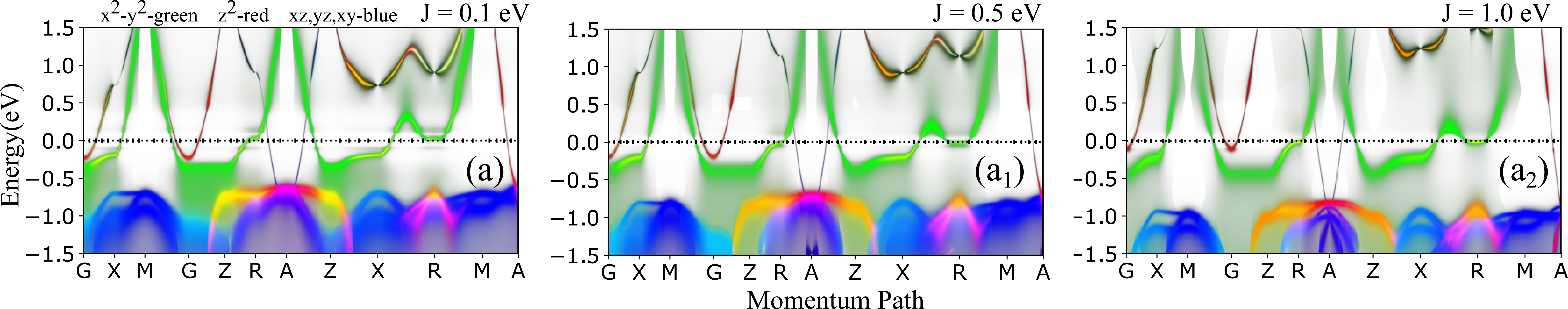}
\caption{(Color online)  \textbf{Orbital projected Spectral functions for LaNiO$_2$} at  U = 5 eV and various J's  in panels (a), (a$_1$) and (a$_2$).}
\label{Hund}
\end{figure*}

\vspace {1.0in}

\section{Supplementary 10: Hybridization of ${\bf Ni$-$d}$ with ${\bf O$-$p}$ and ${\bf La$-$d}$}

\begin{figure*}[!htb]
\includegraphics[width=0.98\linewidth]{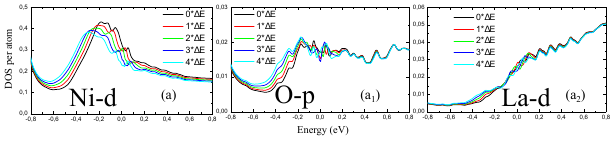}
\caption{(Color online)  \textbf{Density of states (DOS) for LaNiO$_2$:} Panels  (a), (a$_1$) and (a$_2$) shows the dependence of the Ni-d, O-p and La-d projected DOS for the case when the Ni-d DOS is shifted by various amounts of energy $\Delta$E. We observe that when the Ni-d DOS are shifted, the O-p DOS follows more the shift of the Ni-d DOS then does the La-d DOS, implying a stronger hybridization between Ni-d and O-p than 
between Ni-d and La-d.}
\label{Hybri_pd}
\end{figure*}

\newpage

\section{Supplementary 11: Magnetism}

\begin{figure*}[!htb]
\includegraphics[width=0.7\linewidth]{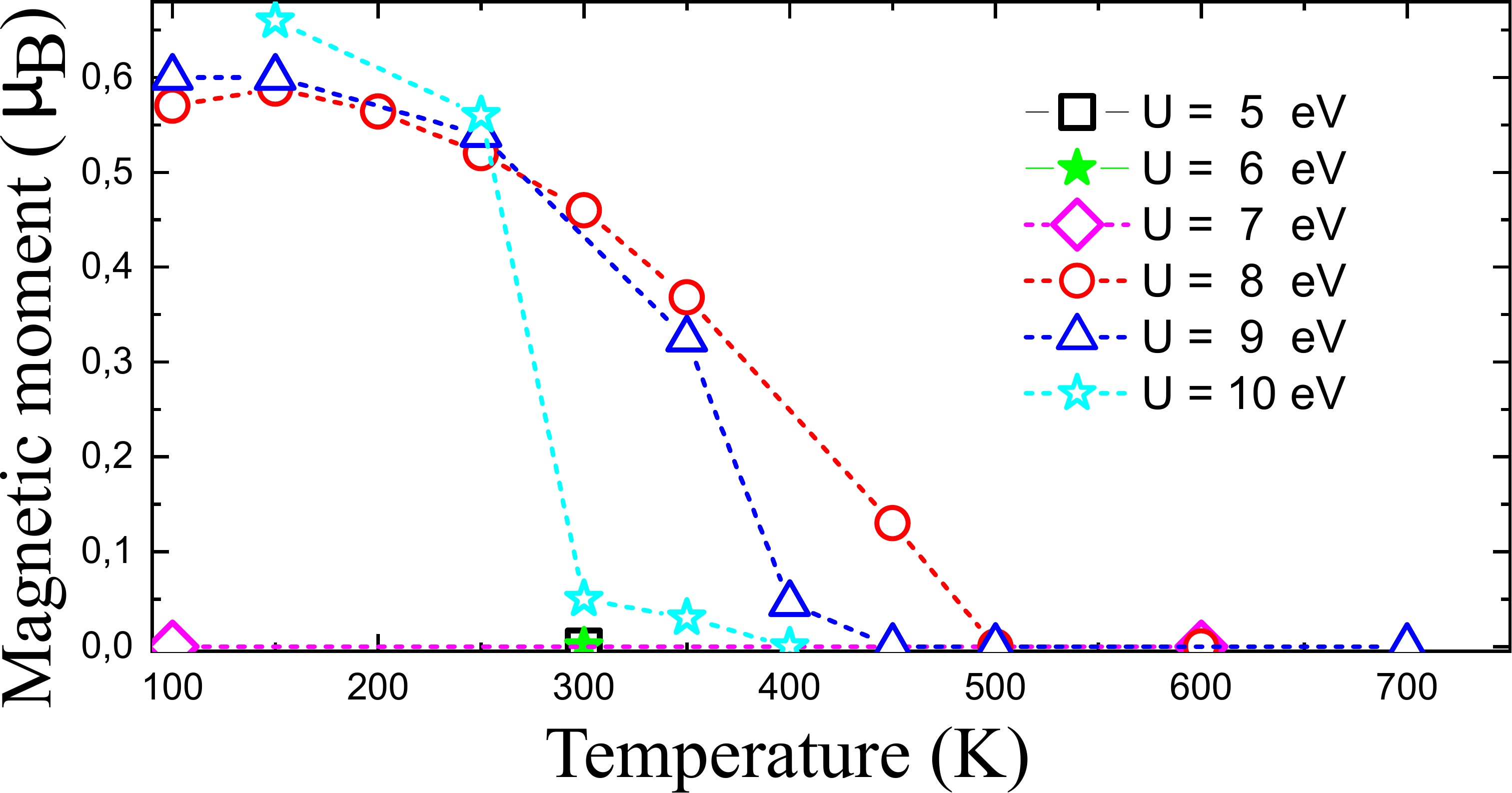}
\caption{(Color online)  \textbf{Order parameter (magnetic moment) for Ni-d ions for the AFM state of LaNiO$_2$ as a function of temperature}, for various values of U.}
\label{Magnetism_Oder_parameter}
\end{figure*}

\vspace {0.7in}

\section{Supplementary 12: Crystal Structure}

\begin{figure*}[!htb]
\includegraphics[width=0.5\linewidth]{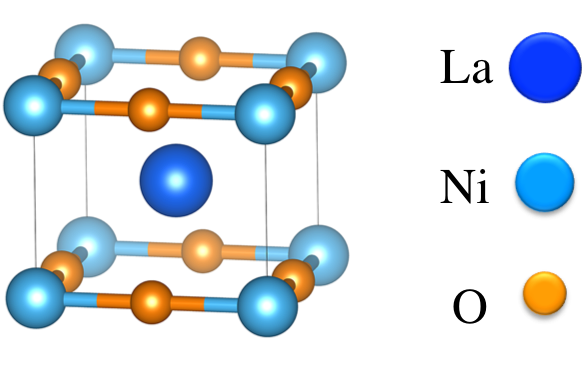}
\caption{(Color online)  \textbf{Tetragonal crystal structure for RNiO$_2$} showing the locations of the rare-earth La, Ni and O elements.}
\label{Crystal_structure}
\end{figure*}
